\documentclass[10pt, conference]{IEEEtran}
\IEEEoverridecommandlockouts

\newcommand{\smallsection}[1]{\noindent\textbf{#1.}}

\makeatletter

\newcommand{\figref}[1]{Figure~\ref{#1}}

\newcommand{\newtextcolor}{blue}
\newcommand{\oldtextcolor}{red}

\newcommand{\todo}[1]{
    \ifthenelse{\equal{#1}{}}{\textcolor{red}{TODO}}{\textcolor{red}{TODO:~{#1}}}
}
\newcommand{\rem}[1]{%
    \ifthenelse{\boolean{showannotations}}%
    {\textcolor{\oldtextcolor}{\st{#1}}}%
    {}%
}

\newcommand\add[1]{%
    \ifthenelse{\boolean{showannotations}}%
    {\textcolor{\newtextcolor}{{#1}}}%
    {#1}%
}

\newcommand\rep[2]{%
    \ifthenelse{\boolean{showannotations}}%
    {\rem{#1}~\add{#2}}%
    {#2}%
}

\newcommand{\openai}{OpenAI\xspace}
\newcommand{\gpt}{GPT-4o\xspace}
\newcommand{\vul}{vulnerability-indicating issues\xspace}

\usepackage{url}

\usepackage{enumerate} 
\usepackage{flushend} 
\usepackage{booktabs} 
\usepackage{color} 
\usepackage{caption}
\usepackage{ifthen}
\usepackage{soul}
\usepackage[normalem]{ulem}
\usepackage{orcidlink}
\usepackage{balance}
\usepackage{eqparbox,collcell}
\usepackage{nicefrac}
\usepackage[export]{adjustbox}
\usepackage[most]{tcolorbox}
\usepackage[export]{adjustbox}
\usepackage{lipsum}
\usepackage{mdframed}
\usepackage{graphicx}
\usepackage{subcaption}
\usepackage{adjustbox}
\usepackage{titlesec}
\usepackage{makecell}
\usepackage{colortbl}
\usepackage{xspace}
\usepackage{listings}
\usepackage{float}
\renewcommand\thesubsection{\Alph{subsection}.}
\renewcommand\thesubsubsection{\thesubsection\Roman{subsubsection}.}
\titleformat{\subsection}
  {\normalfont\bfseries\itshape\normalsize} 
  {\thesubsection}
  {0.5em}
  {}

\titleformat{\subsubsection}
  {\normalfont\normalsize\itshape}
  {\thesubsubsection}
  {0.5em}
  {}
\titlespacing*{\subsubsection}
{0pt}{1.5ex}{1ex}

\newboolean{showannotations}
\setboolean{showannotations}{true} 

\usepackage{cite}
\usepackage{amsmath,amssymb,amsfonts}
\usepackage{algorithmic}
\usepackage{graphicx}
\usepackage{textcomp}
\usepackage{xcolor}
\def\BibTeX{{\rm B\kern-.05em{\sc i\kern-.025em b}\kern-.08em
    T\kern-.1667em\lower.7ex\hbox{E}\kern-.125emX}}
\begin{document}

\title{Detecting Vulnerabilities from Issue Reports for Internet-of-Things}

\author{\IEEEauthorblockN{1\textsuperscript{st} Sogol Masoumzadeh}
\IEEEauthorblockA{\textit{Electrical and Computer Engineering} \\
\textit{McGill University}\\
Montreal, Canada \\
sogol.masoumzadeh@mail.mcgill.ca\orcidlink{0009-0002-6626-1919}}}

\maketitle
\begin{abstract}
Timely identification of issue reports reflecting software vulnerabilities is crucial, particularly for Internet-of-Things (IoT) where analysis is slower than non-IoT systems. While Machine Learning (ML) and Large Language Models (LLMs) detect vulnerability-indicating issues in non-IoT systems, their IoT use remains unexplored. We are the first to tackle this problem by proposing two approaches: (1) combining ML and LLMs with Natural Language Processing (NLP) techniques to detect vulnerability-indicating issues of 21 Eclipse IoT projects and (2) fine-tuning a pre-trained BERT Masked Language Model (MLM) on 11,000 GitHub issues for classifying \vul. Our best performance belongs to a Support Vector Machine (SVM) trained on BERT NLP features, achieving an Area Under the receiver operator characteristic Curve (AUC) of 0.65. The fine-tuned BERT achieves 0.26 accuracy, emphasizing the importance of exposing all data during training. Our contributions set the stage for accurately detecting IoT vulnerabilities from issue reports, similar to non-IoT systems.\end{abstract}


\section{Introduction}
\label{sec: introduction}

\smallsection{Motivation} Submissions to software issue trackers describe bugs, desirable features, and enhancement requests that software teams need to prioritize~\cite{mistrik2010collaborative}. Popular projects attract a large volume of issue reports, making it a challenging task for developers to 
prioritize issues based on their risk level.
Consequently, 
severe issues such as software vulnerabilities may not be prioritized,
endangering dependent products~\cite{dumeric2014matter}. This problem is exacerbated in 
\textit{Internet-of-Things (IoT)}. IoT systems are heterogeneous ecosystems that are composed of loosely interconnected hardware, middleware, and software nodes~\cite{li2015internet}.
Compared to other software, code bases in IoT systems are
more prone to cross-cutting defects~\cite{khezemi2024comparison}. In turn, issues reflecting vulnerabilities are vague and verbose, prolonging their reasoning even more~\cite{memfault2024state}. 
Hence, automated identification of such issues that report on 
vulnerabilities 
in IoT systems (i.e., \textit{\vul}) is of great value. 

\smallsection{Problem} Conventional \textit{Machine Learning (ML)} models~\cite{kallis2019tickettagger} and autoregressive \textit{Large Language Models (LLMs)}~\cite{min2023gptsurvey} are explored for automatically classifying \vul 
in web browsers, cloud, and databases~\cite{behl2014bug, peters2017text, zou2018automatically, das2018security, mostafa2019sais}.
Meanwhile, the application of these models for detecting IoT \vul 
is yet to be investigated.\\
\smallsection{Background and related work} 
Peters et al.~\cite{peters2017text} filtered and ranked $45,940$ issues of Chromium, Wicket, Ambari, Camel, and Derby
to remove non-security issues with security-related tags for enhancing classification performances of text-based models. Das et al.~\cite{das2018security} used ML models and probabilistic \textit{Natural Language Processing (NLP)} techniques to classify issues from~\cite{peters2017text} as security or non-security related. 


\section{Approach and Uniqueness}
\label{sec:methodology}
\begin{figure*}[t]
    \centering
    \scalebox{0.95}{\includegraphics[width=\linewidth]{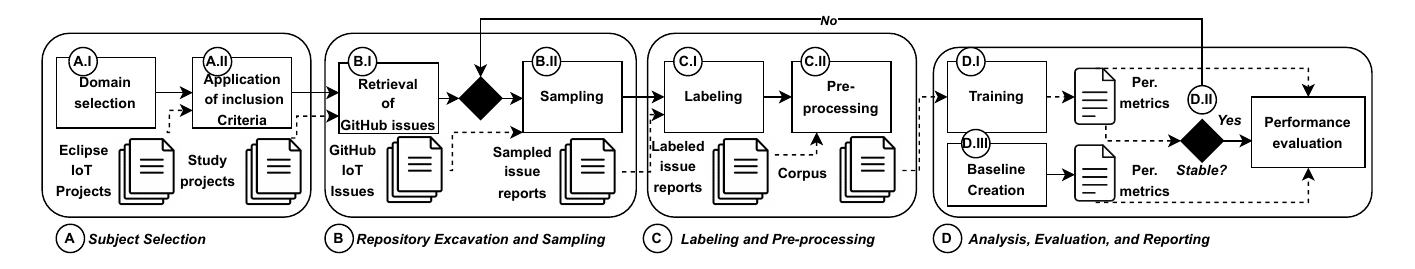}}
    \caption{The overview of the study}
    \label{fig:workflow}
\end{figure*}

We are the \textbf{first, initiating the detection of vulnerabilities in IoT systems from their issue reports}. We propose an empirical study, illustrated in~\figref{fig:workflow}, in which we evaluate the performance of ML models that leverage NLP techniques ($16$ combinations), and \openai{}'s \textit{gpt-4o-1106-preview (\gpt)}~\cite{team2022chatgpt} to detect vulnerability implications in a corpus of issues from \textit{Eclipse IoT} open-source projects. Additionally, we fine-tune a pre-trained neural \textit{Masked Language Model (MLM)}
on an extensive corpus of unlabeled GitHub issues~\cite{meng2020text}, on the downstream task of detecting \vul.

\subsection{The empirical study}
\label{sec:empirical-study}
\begin{figure}[tb]
    \centering
    \includegraphics[width=0.6\linewidth]{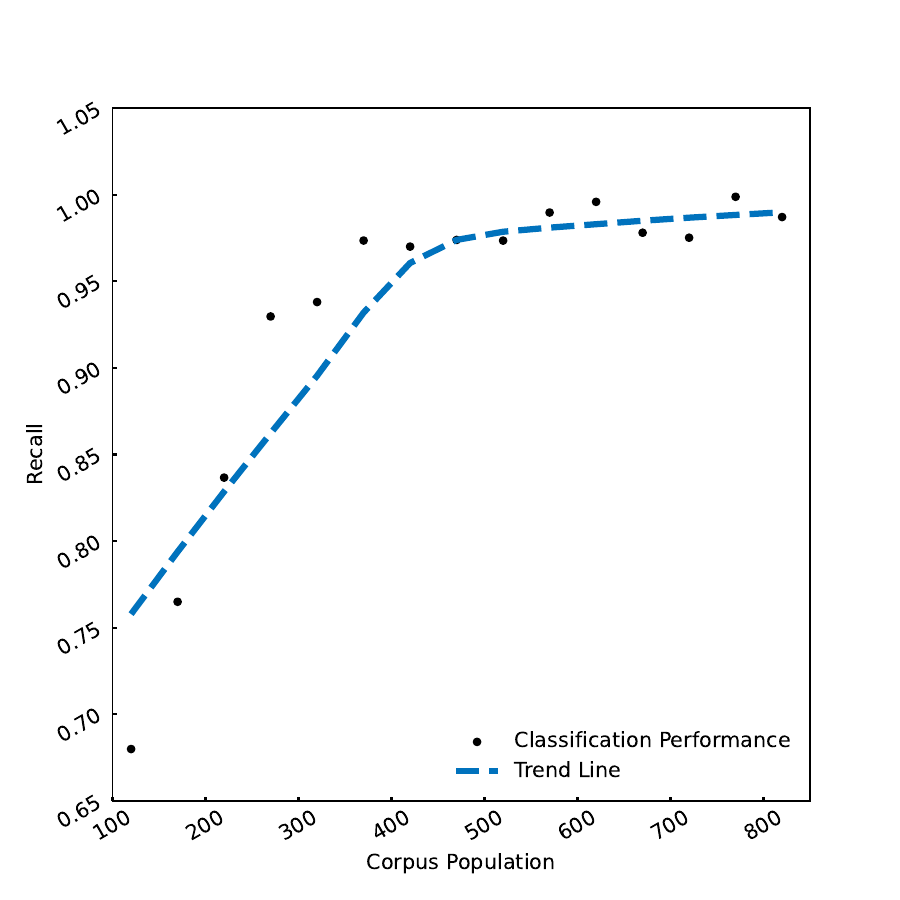}
    \caption{The saturation of the classification performance}
    \label{fig:stability_fig}
\end{figure}
\smallsection{Subjects} From Eclipse IoT projects, we consider active projects with executable CI/CD pipeline and integrated static or dynamic analyzers. $21$ projects satisfy our inclusion criteria. \\
\smallsection{Corpus} We mine GitHub resolved issues of the $21$ Eclipse IoT projects (i.e., $9,564$ reports). From these issues, we randomly sample $370$, providing $95$\% Confidence Level (CL) and $\pm$5\% Confidence Interval (CI). We then expand the corpus by applying stratified sampling, adding $50$ new issues each round, until the classification performance saturates (~\autoref{fig:stability_fig}) at $820$ issues, with $63$ as vulnerability-indicating.\\
\smallsection{Labeling} We label the issues as either vulnerability-indicating or non-vulnerability-indicating based on whether they hint at an exploitable vulnerability in an IoT context. 
The labeling is carried out by three inspectors until 
consensus.\\ 
\smallsection{Pre-processing}\label{sec:pre-processing} 
We pre-process the corpus to remove special characters, English stop words, and memory traces, replace execution logs with semantically equivalent natural text,
and lemmatize inflicted terms.\\
\smallsection{Training} We evaluate the performance of \textit{Gaussian Na\"{\i}ve Bayes (GNB)}~\cite{rish2001empirical}, \textit{Support Vector Machine (SVM)}~\cite{furey2000support}, \textit{Random Forest (RF)}~\cite{biau2016random}, and \textit{Logistic Regression (LR)}~\cite{wright1995logistic} in classifying \vul through \textit{10-fold cross-validation (10-CV)}. We follow the \textit{grid search algorithm} to optimize the classifier hyper-parameters. 
We vectorize our corpus using \textit{BoW}~\cite{wisam2019overview}, \textit{BERT}~\cite{devlin2019bert}, \textit{GloVe}~\cite{pennington2014glove}, and \textit{W2V}~\cite{mikolov2013efficient} NLP features. To balance out the quantities of data classes,
we re-sample the corpus by increasing the population of
\vul. 
Additionally, after carefully handcrafting the prompts, we instruct \gpt to classify \vul.
To account for the LLM's response stochasticity, we set the \textit{Seed} decoding parameter as a constant value. Additionally, we measure the model's performance using \textit{pass@3} which is the upmost correct responses given three generation attempts per issue~\cite{chen2021evaluating}.\\
\smallsection{Evaluation} We evaluate the performance of our models against a \textit{random guesser} with the \textit{Area Under the receiver operator characteristic Curve (AUC)} of $0.50$.
\subsection{The fine-tuning experiment}
\label{sec:learning-experiment}
\smallsection{Corpus} We mine GitHub resolved issues if they are not a pull request, if both title and submission description are non-empty, and if their submission date is between 2022-01-01 and 2024-03-01. From issues satisfying these criteria, those tagged as ``security'' are included as vulnerability-indicating ($6,696$ issues). Issues tagged as ``bug'' but not as ``security'' are included as non-vulnerability-indicating ($528,494$ issues). 
From these issues, we randomly sample $11,000$ with stratification ($95$\% CL and $\pm$5\% CI) as our corpus.\\
\smallsection{The Surrogates} For each issue label, 
we retrieve a list of semantically similar, substitution keywords which we call the \textit{Surrogates}. To do so, (1), we pre-process the corpus by performing the same steps discussed in Section~\ref{sec:empirical-study} Afterwards (2), for each label, we run the \textit{Rapid Automatic Keyword Extraction (RAKE)}~\cite{rose2010automatic} algorithm to determine and rank keywords 
by calculating
frequencies of words in each category and 
analyzing 
their co-occurrences with the rest of the context.
Then (3), for each label,
we manually evaluate the top-$100$ keywords to remove those irrelevant to security-related topics. Finally (4), we discard keywords that appear for both labels.
We set the Surrogates 
as the top-$10$ 
final 
refined keywords for each label.
The Surrogates for vulnerability-indicating and non-vulnerability-indicating labels are: [\textit{CVE}, \textit{GitHub}, \textit{version}, \textit{use}, \textit{vulnerability}, \textit{issue}, \textit{security}, \textit{severity}, \textit{NVD}, and \textit{check}] and [\textit{data}, \textit{file}, \textit{foundry}, \textit{type}, \textit{filter}, \textit{fail}, \textit{error}, \textit{debug}, \textit{default}, and \textit{warn}], respectively.\\
\smallsection{MLM training} We fine-tune a 
BERT MLM,
hiding 
occurrences of the Surrogates in the corpus, using the \textsf{[MASK]} token. 10-CV, with training epochs=2 and batch size=5, is executed on a NVIDIA T4 hardware accelerator.
\begin{table}[H]
\caption{The empirical study results \begin{small}(AUC values)\end{small}}
\label{tab:empirical-study-results}
\centering
\footnotesize

\begin{tabular}{@{}crrrr@{}}
\toprule
             & \multicolumn{1}{l}{\textbf{BoW}} & \multicolumn{1}{l}{\textbf{BERT}} & \multicolumn{1}{l}{\textbf{GloVE}} & \multicolumn{1}{l}{\textbf{W2V}} \\ \cmidrule(l){2-5} 
\textbf{GNB} & 0.53                             & 0.62                              & 0.54                               & 0.55                             \\
\textbf{SVM} & 0.55                             & \textbf{0.65}                     & 0.61                               & 0.58                             \\
\textbf{RF}  & 0.45                             & 0.63                              & 0.58                               & 0.60                             \\
\textbf{LR}  & 0.58                             & 0.59                              & 0.45                               & 0.44                             \\ \bottomrule
\end{tabular}%
\end{table}

\section{Results and Contributions}
\label{sec:results}
Performances of ML models in classifying vulnerability-indicating issues are demonstrated in
Table~\ref{tab:empirical-study-results}. For the same experiment, \gpt achieves an AUC of $0.60$.
Our most accurate classifier is the SVM trained using BERT NLP features
and achieves an AUC of $0.65$. On the other hand, the mean classification accuracy of the fine-tuned BERT MLM is only $0.26$ across ten training folds. This considerable difference between the performances of the classification settings highlights the importance of exposing all the context to the classifiers while training. Although the corpus of the fine-tuning experiment is much more extensive compared to the corpus of the empirical study, masking the Surrogates prevents the model to learn the entire data: (1) many issues do not contain any of the Surrogates and hence are not exposed to BERT MLM
and (2) the model is not trained on the \textsf{[CLS]} token to predict the label while seeing the whole issue description.\\ \smallsection{Future work} We plan to add an unsupervised training layer that exposes BERT MLM to all issues, allowing it to iteratively refine its prediction generalizability. We also aim to extend the experiments with additional baselines and larger datasets.
\flushend
\IEEEtriggeratref{12}

\bibliographystyle{IEEEtran}
\bibliography{99_references}
\end{document}